# Chiral Spontaneous Photon-Pair Generation with Plasmonic Bound States in the Continuum

SKY SEMONE,[1] MATTHEW J. BRANDSEMA,[2] CHRISTOS ARGYROPOULOS[1,*]

[1] Department of Electrical Engineering, The Pennsylvania State University, University Park, PA 16803
[2] Applied Research Labs, The Pennsylvania State University, University Park, PA 16803
*cfa5361@psu.edu

**Spontaneous parametric down conversion (SPDC) has proven to be a robust and prominent method for creating non-classical light sources of entangled single-photon pairs. However, such sources suffer from low efficiency due to the inherent weakness of the SPDC process. Moreover, there is no control of polarization in the generated photons. Circular polarized SPDC-based sources with efficient chiral response are currently missing, which severely precludes the realization of a wide range of practical applications in quantum computing and communications. The current work solves these inherent problems by demonstrating a novel chiral plasmonic metasurface that leverages a strong quasi-bound state in the continuum (qBIC) resonance which significantly increases the SPDC photon generation efficiency and realizes circular polarized (chiral) single-photon pairs. The presented metasurface design is ultrathin, supports efficient free space outcoupling of chiral classical and quantum radiation, and operates at room temperature. Hence, it offers the potential to realize a compact and integrable quantum source of circularly polarized entangled single-photon pairs.**

Quantum photonics is a rapidly growing field with promising applications in quantum communications, sensing, and computation [1]. At the core of much of its ongoing technology are quantum sources that generate entangled and correlated single-photon pairs. The most commonly used method for producing these photon pairs is spontaneous parametric down-conversion (SPDC) [2], which, although vital to many quantum technologies, has limitations of very low efficiency due to the inherently weak nature of quantum nonlinear interactions combined with lack of polarization control. Enhancing the efficiency of single-photon non-classical sources and enabling precise polarization control of quantum light are major emerging research avenues, as existing quantum sources present limitations and technological hurdles that hinder their development and practical deployment in quantum devices.

Contemporary designs of SPDC sources often utilize nonlinear crystals, nanowaveguides, and ring resonators to improve performance [2]; however, these approaches introduce design limitations that complicate practical integration. Moreover, recently, dielectric and metallic (plasmonic) metasurfaces have successfully engineered the spatial and spectral correlations of single-photon pairs generated by SPDC [3–10]. However, none of these designs support circularly polarized (chiral) single-photon emission, which is crucial for realizing photonic quantum bits, also known as flying qubits [1]. In this work, we solve this problem by presenting nanoscale metallic (plasmonic) metasurfaces design that significantly enhance single-photon generation while at the same time, enabling preferential chiral quantum light emission. The proposed compact quantum light source design can be easily coupled to free space and operate at room temperature, eliminating the need for stringent thermal control—making it well-suited for real-world applications such as quantum LiDAR and sensing technologies [11].

The metasurface consists of a silver asymmetric grating structure composed of periodic metallic nanorods placed at an oblique angle nearby a longer but smaller width continuous nanobar along the y-direction, as displayed in Fig. 1(a). Such a metasurface design can achieve high quality-factor (Q-factor) chiroptical resonances that enhance chiral light-matter interactions by breaking both in-plane inversion and mirror symmetries, resulting in a chiral quasi-bound state in the continuum (qBIC) narrowband response. Unlike dielectric metasurfaces which have qBIC responses that usually cannot achieve substantially enhanced fields, metallic (plasmonic) configurations like the one leveraged in this research can deliver a large electric field enhancement in the dielectric layer [12,13]. Only recently have a few qBIC metasurface designs using plasmonic materials been proposed [14–16]. However, all existing dielectric and plasmonic qBIC metasurfaces to date have been achiral, i.e., do not interact preferentially to circular polarized incident radiation and do not emit chiral photons. To the best of our knowledge, the presented design of Fig. 1 is the first demonstration of a chiral plasmonic qBIC nonlinear and quantum optical metasurface. It combines strong optical chirality with a high-Q resonance and significant field enhancement within the underlying dielectric layer—making it an excellent candidate for enhancing nonlinear and quantum optical phenomena.

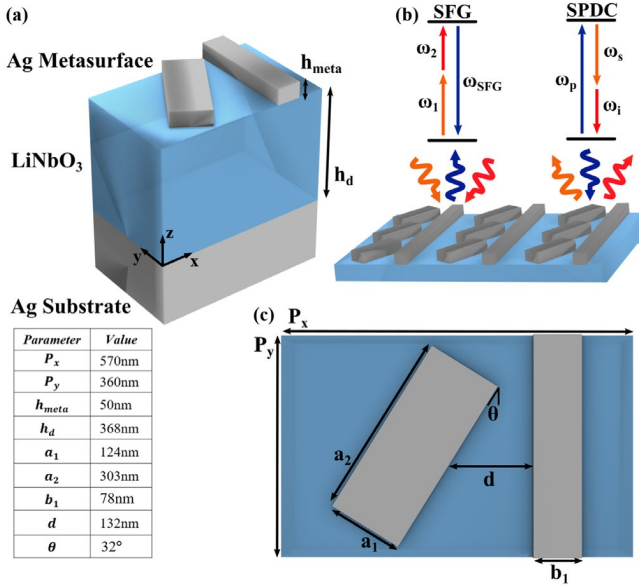

Fig. 1. (a) Chiral q-BIC plasmonic metasurface unit cell. (b) Classical nonlinear SFG and quantum SPDC processes. (c) Dimensions of metasurface unit cell.

Here, we focus our attention on enhancing SPDC-based entangled single-photon pair emission. To that end, we select x-cut lithium niobate (LiNbO$_3$) as the dielectric spacer layer due to its strong second-order nonlinear susceptibility, which is essential to achieve efficient sum-frequency generation (SFG)—a classical nonlinear process that, through quantum-classical correspondence, can enhance SPDC efficiency [10]. To induce chirality, one of the two asymmetric silver nanorods is tilted in-plane, breaking the in-plane and mirror symmetry and enabling a chiral qBIC optical response. Additionally, the underlying silver substrate supports a Fabry–Perot mode confined beneath the metasurface, within the nonlinear dielectric layer, effectively forming an extremely subwavelength nanocavity. This configuration excites localized surface plasmon resonances (due to nonlocality), and gap plasmon resonances (due to the formed nanocavity), which results in highly enhanced electric fields in the LiNbO$_3$ layer. Since the nonlinear optical response scales with the electric field intensity, the metasurface substantially improves the nonlinear conversion efficiency compared to bulk LiNbO$_3$ alone. Importantly, the nanorods asymmetric widths realize a qBIC response, producing sharp, high-Q resonances that contribute to both strong field enhancement and spectral selectivity. The in-plane rotation of the nanorods not only drives chirality but also yields a polarization-selective enhancement in field intensity and single-photon emission, favoring either left-handed (LH) or right-handed (RH) circular polarized flying qubits. Note that although this study uses a specific parameter set to realize the presented optimized metasurface, the proposed design is highly tunable and adaptable to different materials and spectral regimes just by varying its dimensions and properties. As such, chiral qBIC plasmonic metasurfaces are promising candidates for next-generation integrated quantum light sources, particularly in generating circularly polarized entangled single-photon qubits for advanced quantum communication and sensing technologies.

The metasurface unit cell is depicted in Fig. 1(c) along with the complete set of geometric parameters. The foundational silver layer acts as an opaque substrate that prevents light from transmitting through the structure and is set to 100 nm, significantly thicker than the metal skin depth to eliminate potential transmission. It also acts as an interface to support plasmonic resonances formed at the LiNbO$_3$ layer nanogap below the metasurface. To emulate naturally occurring imperfections due to the lithography process in our simulations, a 2 nm radius curvature is introduced to all nanorod edges. Fabrication constraints reflective of practical direct-write lithography limitations are incorporated into the optimization process, ensuring the design remains feasible for real-world implementation. Additionally, the thickness of the LiNbO$_3$ layer in the optimized qBIC metasurface design is chosen to be 368nm, which has been fabricated over a metallic substrate in previous research [17]. The physical parameters given here result from several rounds of optimization, where the ultimate objective function simultaneously maximizes the electric field enhancement for LH circular polarization while minimizing it for RH polarization.

The linear chiroptical response of the qBIC metasurface is evaluated by illuminating it from top with circularly polarized light and analyzing both the absorption and the electric field enhancement, quantified as $|E/E_0|$, where $E$ is the local electric field and $E_0$ is the amplitude of the incident field. Complete details of the simulations and material properties used are provided in Supplement 1. Figure 2(a) presents the absorption spectra for LH and RH circularly polarized light, as well as the circular dichroism (CD) in the inset, defined as the difference in absorption between the two circular polarizations: $CD = A_{LH} - A_{RH}$. The absorption plot reveals a pronounced resonance at $\lambda = 1279$ nm, where absorption of LH-polarized light approaches unity while RH absorption remains minimal—resulting in exceptionally high circular dichroism, as shown in the inset plot of Fig. 2(a). A secondary higher-order resonance is observed at $\lambda = 746$ nm, with corresponding peaks in both absorption and CD. This is the main reason why we picked these two resonances in all our calculations, since at these wavelengths the highest chiral response is expected combined with the strongest electric field enhancement shown in Fig. 2(b). Figures 2(b) and 2(c) display the electric field enhancement distributions in the x- and y-planes at both resonance wavelengths (left side/ $\lambda = 746$ nm and right side/ $\lambda = 1279$ nm) for LH-polarization (Fig. 2(b)) and RH-polarization (Fig. 2(c)). Notably, LH-polarized excitation yields significantly stronger field confinement and enhancement in the nanogap and along the asymmetric metasurface compared to RH-polarized, indicating a robust chiral coupling of the localized optical response. Finally, the induced magnetic field when the metasurface is excited by LH-polarized incident wave is shown in Figs. 3(a) and 3(b) at the two resonant wavelengths of $\lambda = 746$ nm and $\lambda = 1279$ nm, respectively, where arrows are used to demonstrate the electric field circulating direction. From these magnetic field results, it is clear that the $\lambda = 1279$ nm fundamental resonance is due to a magnetic dipole resonance, similar to other nanogap metasurfaces [10], while the $\lambda = 746$ nm resonance is higher-order and quadrupolar in nature.

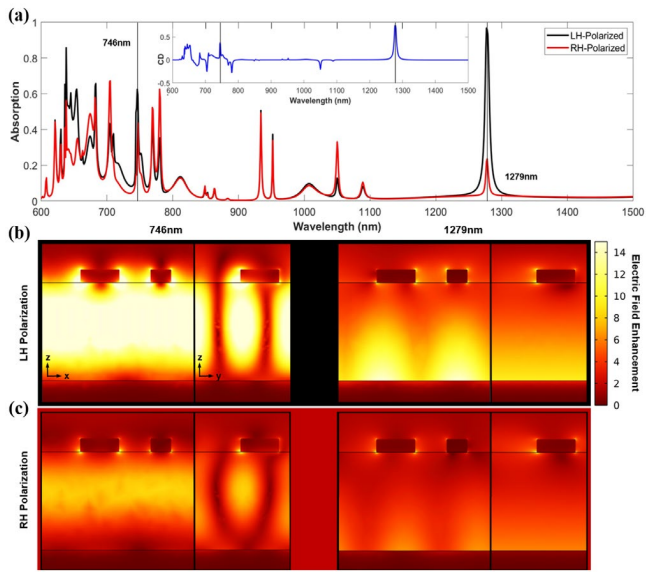

Fig. 2. (a) Absorption caused by the metasurface for LH- and RH-circular polarized incident light leading to high circular dichroism (shown in the inset) at wavelengths 746 nm and 1279 nm. (b)-(c) At each wavelength, the strong chiral absorption enhances the electric field significantly in the nanogap and along the asymmetric width nanostrips for (b) LH incident wave polarization compared to (c) RH incident polarization.

While ideal BICs are a mathematical abstraction characterized by infinitely high-Q resonances, practical metasurface designs can achieve quasi-BIC behavior by introducing controlled asymmetries, yielding high, but finite Q-factors [13]. Such qBIC metasurfaces exhibit sharply defined resonances that can be further narrowed until material losses become the dominant limiting factor in the resulted narrow bandwidth. Recently, chiral plasmonic metasurfaces have been shown to display qBIC resonances [18], however, with lower chiroptical response mainly due to not using the currently proposed asymmetric width nanorods that can further enhance the localized field and spectral selectivity (see Supplement 1 for qBIC metasurface further discussion).

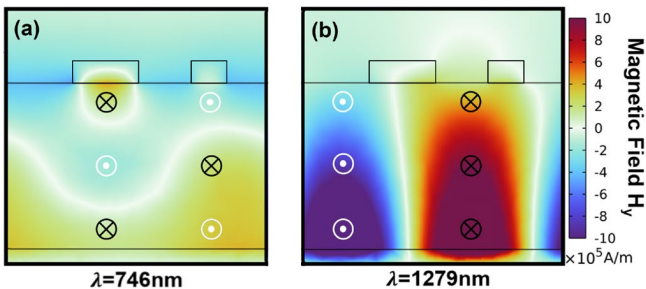

Fig. 3. Normalized y-component of the magnetic field for the two resonant wavelengths (a) λ=746 nm and (b) λ=1279 nm. The arrows are used to demonstrate the electric field direction.

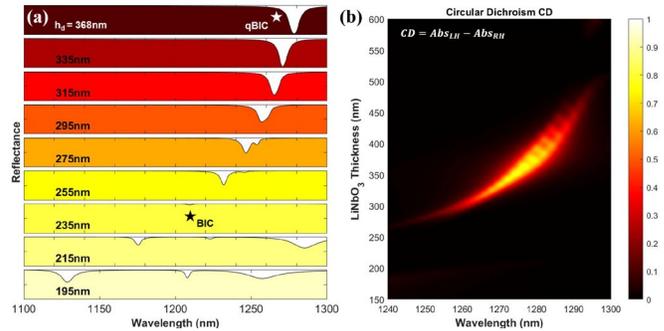

Fig. 4. (a) Quasi-BIC behavior is demonstrated by the high Q-factor resonances in the reflectance as LiNbO$_3$ thickness h$_d$ varies from 195nm to 368nm for LH-polarized light. When the LiNbO$_3$ thickness is 235nm, almost perfect BIC (star sign) is realized as the Q-factor nears infinity and the corresponding dip in reflectance is minimal. (b) When the LiNbO$_3$ thickness is fixed to 368nm, the qBIC geometry is highly absorbative to incident light under LH-polarized illumination, while reflecting RH-polarized illumination, causing a maximum circular dichroism of 0.8.

In metal–dielectric–metal metasurfaces, like the one employed in our design, this qBIC behavior can be tuned by varying structural parameters such as the illumination angle or the dielectric gap thickness [14]. In our study, we maintain varying structural parameters such as the illumination angle to normal incidence and systematically vary the gap thickness to demonstrate the increasing Q-factor. These trends are clearly illustrated in Fig. 4. In the ideal BIC regime—where the Q-factor approaches infinity—radiative losses vanish, and as a result, the structure exhibits perfect reflection, like a plain metallic substrate, as shown in Fig. 4(a) for LiNbO$_3$ layer thickness of 235nm. However, varying the nanogap thickness reduces the Q-factor and gives rise to radiative qBIC modes, which in turn produce strong chiral absorption and significant electric field enhancement, leading to pronounced dips in reflectance, as depicted in Fig. 4(a).

When the LiNbO$_3$ layer thickness is 368 nm, our structure reaches peak absorption for LH circularly polarized light due to critical coupling with incident chiral radiation. Such thickness value yields the strongest field confinement and, correspondingly, the highest circular dichroism, as illustrated in Fig. 4(b), which maps CD as a function of wavelength and dielectric nanogap thickness. Therefore, we select to further explore this geometry at the optimal operation point, where enhanced chirality and boosted field within the nanogap layer are simultaneously achieved.

The next step is to investigate the nonlinear response of our metasurface. More specifically, we study SFG which is a nonlinear process where two pump photon beams $\omega_1$ and $\omega_2$ undergo mixing in a nonlinear material to produce a third photon beam with frequency $\omega_{SFG} = \omega_1 + \omega_2$. It can be considered as the classical inverse counterpart of non-degenerate quantum SPDC process, where a pump photon $\omega_p$ transforms into a set of signal and idler entangled single photons with conserved energy $\omega_p = \omega_s + \omega_i$ [10]. Both these processes are schematically depicted in Fig. 1(b).

To avoid the complexities of the statistical nature of quantum simulations, the quantum-classical correspondence principle is used to relate the SPDC process, that we wish to investigate, to the nonlinear optical SFG model which can be simulated by a classical nonlinear electromagnetic solver, however substantially revised by introducing nonlinear polarizabilities and two incident waves, as was shown in previous work [10]. Specifically, the following formula determines the generation rate of single-photon pairs per second as a function of the classical SFG efficiency $\eta_{SFG}$: $\frac{dN_{pair}}{dt} = \pi c \frac{\lambda_p^4}{\lambda_s^5 \lambda_i^3} \frac{I_p \Delta \lambda_s}{A} \eta_{SFG}$, where $\lambda_p$, $\lambda_s$, and $\lambda_i$ are the pump, signal, and idler photons, respectively, $c$ is the speed of light in vacuum, $I_p$ is the pump intensity, $A$ is the area of illumination along the metasurface, and $\Delta \lambda_s$ is the spectral bandwidth of the illuminated signal [10]. The spectral bandwidth is governed by both the response characteristics of the metasurface, and the sensitivity range of the detectors used and is set to 1nm for this study which is typically used value [10]. The SFG efficiency is defined as the total power of the SFG wave divided by the intensities of both incident waves. Note that the pump intensity is kept relatively low ($I_p = 5mW$), since high intensity values can lead to detrimental stimulated photon generation effects or nonlinear saturated absorption that can cause a decrease in the efficiency of the SPDC or SFG processes, respectively [2].

Simulations of the SFG process were performed using circular (LH, RH) and linear (X, Y) polarizations for incident waves always fixed to 1279 nm and 746 nm, with the generated SFG power recorded to compute conversion efficiency for each polarization combination. The relevant results are shown in the histogram of Fig. 5(a). Because the metasurface was designed to have a high enhancement only in the case of LH-polarization and a low enhancement for RH-polarization, the resulting efficiency follows this trend as well, as can be seen by the star and triangle signs in Fig. 5(a), respectively. The nonlinear simulations reveal that the highest SFG efficiency occurs when both incident waves are LH-polarized, while those with at least one RH-polarized signal yield significantly lower efficiency. The strong distinction between LH-LH and RH-RH polarizations highlights the pronounced chiral behavior of the metasurface but now extended to the SFG nonlinear regime. Interestingly, cross-linear polarized waves (X and Y and the inverse) or circular (LH) to linear (X or Y) can also lead to high SFG efficiency, however, with lower values compared to the purely chiral response. This is expected since the metasurface has strong resonant response even in the case of linear polarized illumination. As a final step, the photon-pair generation rate via SPDC is calculated for both LH- and RH-polarized entangled single-photon pair emission. The results, shown in Fig. 5(b), indicate that the metasurface can generate up to 110,000 photon pairs per second in the case of LH-polarized emission, while the RH-polarized emission is significantly lower at 4,800 photon pairs per second. This demonstrates the metasurface strong quantum chiral light emission capabilities.

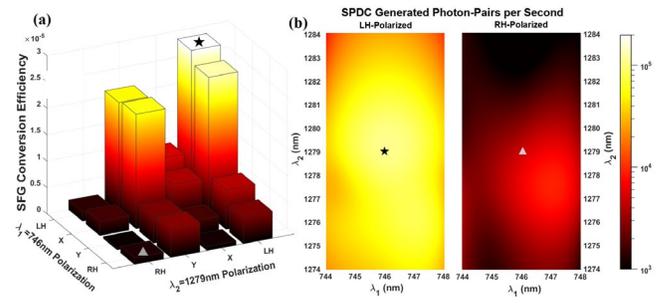

Fig. 5. (a) Computed SFG conversion efficiency at the two qBIC metasurface resonances 1279 nm and 746 nm for various polarization illuminations. (b) SPDC generation rates for LH- and RH-polarized single-photon emission.

In conclusion, we demonstrate a novel and compact chiral entangled single-photon pair source based on a qBIC plasmonic metasurface with nanoscale features, operating effectively at room temperature. In addition to the chiroptical single-photon emission, the currently realized metasurface has record-breaking high SPDC efficiency compared to relevant existing literature (see Supplement 1, Table S1). The resulting strongly chiral and intense quantum light emission offers a promising platform for the realization of photonic qubits, enabling advanced applications in next-generation quantum communication and sensing technologies

**Funding.** Penn State Applied Research Laboratory Walker Fellowship, NSF 2224456.

**Disclosures**. The authors declare no conflict of interest.

**Data Availability Statement (DAS).** Data may be obtained from the authors upon reasonable request.

**Supplemental Document.** See Supplement 1.

# SUPPLEMENT 1

## CHIRAL SPONTANEOUS PHOTON-PAIR GENERATION WITH PLASMONIC BOUND STATES IN THE CONTINUUM: SUPPLEMENTAL DOCUMENT

SKY SEMONE,[1] MATTHEW J. BRANDSEMA,[2] CHRISTOS ARGYROPOULOS[1,*]

[1] *Department of Electrical Engineering, The Pennsylvania State University, University Park, PA 168032*
[2] *Applied Research Lab, The Pennsylvania State University, University Park, PA 16803*
*[*cfa5361@psu.edu](mailto:cfa5361@psu.edu)*

## 1. Material Properties and Simulations Details

Figure 1 in the main paper shows one unit cell of the proposed metasurface geometry, consisting of a periodic array of silver nanostripes positioned atop a dielectric spacer layer composed of lithium niobate (LiNbO$_3$). A continuous silver substrate beneath the spacer serves as a reflective layer, enabling the metasurface to operate in reflective mode. This configuration significantly enhances the electric field within the LiNbO$_3$ layer, as demonstrated in Fig. 2(b). The oblique asymmetric width nanobars require the use of 3D simulations to be accurately modeled. In the simulation domain, two ports of circular polarization are placed at each top and bottom boundaries to support excitation for both incident wavelengths, λ$_1$ and λ$_2$. Owing to the structure's periodicity, a single unit cell containing one straight and one oblique silver nanostripe is simulated, with periodic boundary conditions applied to the lateral (left and right) edges. The incident radiation is kept at normal angle to the surface for all simulations in this study.

LiNbO$_3$ is an emerging and currently widely used bulk nonlinear material due to its relatively strong second-order nonlinearity [1–3]. Its second order nonlinear response can be expressed completely by three orthogonal components of the induced anisotropic nonlinear polarizability in the utilized sum-frequency generation (SFG) process which are given by:

$$P_x^{NL} = 2\varepsilon_0 \left[ d_{33} E_{1_x} E_{2_x} + d_{31} \left( E_{1_y} E_{2_y} + E_{1_z} E_{2_z} \right) \right], \quad (S1)$$

$$P_y^{NL} = 2\varepsilon_0 \left[ d_{31} \left( E_{1_y} E_{2_x} + E_{1_x} E_{2_y} \right) + d_{22} \left( E_{1_y} E_{2_y} - E_{1_z} E_{2_z} \right) \right], \quad (S2)$$

$$P_z^{NL} = 2\varepsilon_0 \left[ d_{31} \left( E_{1_z} E_{2_x} + E_{1_x} E_{2_z} \right) - d_{22} \left( E_{1_y} E_{2_z} + E_{1_z} E_{2_y} \right) \right], \quad (S3)$$

where $\varepsilon_0$ is the permittivity of free space, and $d_{33} = 34.4\,pm\,V^{-1}$, $d_{31} = 5.95\,pm\,V^{-1}$, $d_{22} = 3.07\,pm\,V^{-1}$ are the non-zero elements of the LiNbO$_3$ crystal anisotropic second-order nonlinear susceptibilities [4]. The subscripts x, y, and z in Equations (S1-S3) represent the corresponding components of the induced nonlinear polarizability and electric field along different axes. In order to utilize the predominant $d_{33}$ nonlinear LiNbO$_3$ susceptibility with the plasmonic metasurface described in this study, the LiNbO$_3$ crystal is x-cut, with its optical axis aligned along the x-direction. Finally, it should be noted that the refractive index of silver is interpolated in COMSOL over the range of wavelengths used in this study from experimental data [5], while the values of the LiNbO$_3$ layer were also taken from relevant experimental measured data [6].

In this study, both linear and nonlinear SFG processes are numerically modeled in the frequency domain using COMSOL Multiphysics, a commercial simulation platform based on the finite element method (FEM). The simulations are conducted using the software's RF Module to perform full-wave electromagnetic analyses. These simulations require modification of the default solver equations to incorporate the nonlinear optical response of LiNbO$_3$, which are expressed by the non-zero term $\mu_0 \omega^2 \mathbf{P}^{NL}$ using the following Equation S4:

$$\nabla \times (\mu_r^{-1} \nabla \times \mathbf{E}) - \varepsilon_r k_0^2 \mathbf{E} = \mu_0 \omega^2 \mathbf{P}^{NL} \quad (S4)$$

To express the second-order nonlinear polarization $\mathbf{P}^{NL}$ in COMSOL, weak-form contributions are created for each term given by Equations S1-S3. Similar FEM-based simulation techniques have been extensively employed to model a range of nonlinear optical phenomena, including SFG in plasmonic and dielectric systems, and have been validated through comparison with experimental results [7–9].

## 2. Summary of Single Photon Pair Emission from SPDC Sources

A key component of many quantum technologies is the ability to generate entangled and correlated single-photon pairs which are most commonly created by the spontaneous parametric down-conversion (SPDC) process. While SPDC plays a crucial role in current quantum systems, it suffers from low efficiency due to the inherently weak nature of quantum nonlinear interactions and limited polarization control. **Table S1** highlights many of the recent advances in using thin films and metasurfaces to increase the efficiency of the SPDC single-photon pair generation.

More specifically, in **Table S1**, the thickness column measures the thickness of the nonlinear material used to generate the SPDC photon pairs and is used to evaluate efficiency. Enhancement refers to the enhancement in the number of SPDC photon pairs attributed to the metasurface used in each respective study compared to the same nonlinear material without the metasurface. Resonance types presented in **Table S1** are to provide physical insights in the primary method of SPDC enhancement. The abbreviations used in the table for these resonance types are:
- GMR: Guided-mode resonance
- BIC: Bound states in the continuum
- FP: Fabry-Perot
- LSPR: Localized surface plasmon resonance

Measured rate and enhancement are affected by several external factors, including the bandwidth of the filters, the collection geometry and efficiency, the performance of the single-photon detectors, and the optical components in the system. Consequently, the measured values do not accurately represent the intrinsic photon pair generation rate within the metasurface internally. When experimental data is available, coincident-to-accident ratio (CAR) is presented as reported by the originating research, or by means of the second-order cross-correlation function using the formula: $CAR = g^{(2)}(0) - 1$.

**Table S1. Nanophotonic SPDC Single Photon Pair Sources**

| Material and Geometry | Thickness (um) | Resonance Type | Pump (mW) | Rate (Hz) | Efficiency (Hz/mW/um) | Enhancement | CAR | Type | Sources |
|---|---|---|---|---|---|---|---|---|---|
| $LiNbO_3$ | .680 | Mie | 70 | 5.4 | 0.11 | 20 | 361 | Exp | [10] |
| $LiNbO_3$ | .304 | GMR | 85 | 1.8 | 0.07 | 450 | 5000 | Exp | [11] |
| GaAs | .500 | BIC | 9 | 0.08 | 0.018 | >1000 | 9.5 | Exp | [12] |
| $LiNbO_3$ | .300 | GMR | 85 | 0.83 | 0.033 | 210 | 1700 | Exp | [13] |
| GaP | .150 | BIC | 70 | 0.24 | 0.022 | 67 | 4.8 | Exp | [14] |
| $LiNbO_3$ | .308 | GMR | 91 | 2.29 | 0.082 | N/A | 7500 | Exp | [15] |
| $LiNbO_3$, film | 6.00 | N/A | 220 | 1400 | 1.06 | N/A | 4.8 | Exp | [16] |
| $LiNbO_3$, film | .300 | FP | 9 | 2.81 | 1.04 | ~ | 1000 | Exp | [17] |
| $LiNbO_3$, nanocubes | 3.600 | N/A | 50 | 1.26 | 0.007 | ~ | 360 | Exp | [18] |
| $LiNbO_3$, Ag metasurface | .306 | LSPR+ FP | 5 | 1,400 | 915 | >1000 | ~ | Sim | [19] |
| **$LiNbO_3$, Ag metasurface** | **.368** | **BIC + LSPR** | **5** | **110k** | **60k** | **>50,000** | **~** | **Sim** | **This research** |

## 3. Quasi Bound States in the Continuum Chiral and Achiral Plasmonic Metasurfaces

Bound states in the continuum (BICs) are localized states that, despite residing within the energy range of extended radiative modes (the continuum), remain completely confined and non-radiative. They defy conventional expectations of wave propagation and leakage and have generated intense interest in optics and acoustics. Two major classes of BICs are symmetry-protected BICs and accidental BICs, which can be rigorously distinguished based on their symmetry constraints, parameter sensitivity, and topological characteristics [20,21].

Symmetry-protected BICs arise due to an incompatibility between the symmetry of the localized mode and that of the available radiative continuum channels. In such cases, even though the mode lies within the frequency range of the continuum, it cannot be coupled to the outgoing radiation due to orthogonality enforced by the system's spatial or temporal symmetries. Accidental BICs emerge from destructive interference or fine-tuned interference conditions that suppress radiation in all allowed channels instead of symmetry. They do not require any specific symmetry and can occur generically in systems with broken or low symmetry. These BICs are "accidental" in the sense that they appear only for specific values of continuous system parameters such as geometric dimensions, refractive index contrasts, or material anisotropies, where interference conditions become precisely met.

The BIC and quasi-BIC resonances created by the plasmonic metasurface in this study can be classified as accidental, because they are highly dependent upon the geometrical tuning of the metasurface parameters, the thickness of the underlying oxide, along with the angle and wavelength of the incident light. In Fig. 4(a) in the main text, we show the effects of changing incident wavelength and oxide thickness, where BIC is observed at oxide thickness $h_d = 235nm$ and wavelength $\lambda = 1208nm$. As the thickness of the LiNbO$_3$ layer is either increased or decreased, there is a corresponding wavelength that will be absorbed by the formed quasi-BIC resonance due to the chiral plasmonic metasurface. The progression of BIC resonances into qBIC is shown through the relationship between wavelength and LiNbO$_3$ thickness, as depicted in Fig. 4(a) in the main text. However, such progression can also be seen as the angle of incident light varies when impinging on the metasurface. In Figure S1, the angular dependence of the high-Q qBIC resonance can be seen, notably with a high sensitivity to changes of incident angle in the x-direction.

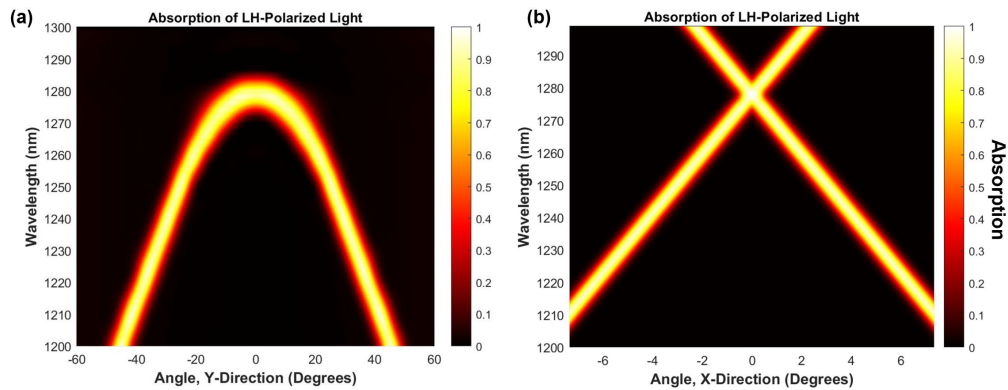

Fig. S1. The absorbance of the metasurface with LH-polarized light as the incident angle is altered from normal incidence in the (a) y-direction and (b) x-direction. The metasurface is particularly sensitive in the x-direction, where only a few degrees can drastically change the spectral response of the resonance.

On a relevant note to the current presented chiroptical nanophotonic system, several sensors based on resonant chiral plasmonic metasurfaces have been developed, such as plasmonic sensors with applications in biomedicine, environmental monitoring, and chemical sensing [22–24]. These show great improvement over naturally found chiral materials that exhibit a much weaker chiral response and reduced sensitivity. Chiral plasmonic metasurfaces working in the infrared regime play a crucial role in chiral thermal switches, selective molecular sensing, and thermophotovoltaics [25–27].

Despite the broad success of chiral metasurfaces, they still face some limitations in the visible and near-infrared bands, such as absorption and scattering losses, discontinuous response, and limited chiroptical response. High Q-factor BIC metasurface designs can enhance local electric field intensity and display extreme sensitivity, useful for a range of applications including sensing [28–30], lasers [31–34], light sources [35,36], and nonlinear fields [37–41].